\documentclass[runningheads]{llncs}

\usepackage{graphicx}
\usepackage{float} 
\usepackage{caption}
\usepackage{cite}
\usepackage{subcaption}
\usepackage[normalem]{ulem} % for strike-through
\usepackage{enumitem}
\usepackage{hyperref}
\usepackage{longtable}
\usepackage{pgfplots}
\pgfplotsset{compat=1.7}
\usepackage{pgf-pie}
\usepackage{xpatch}
\makeatletter
\def\setcolor#1\pgfeov{\def\pgfpie@color{#1}}
\pgfkeyslet{/color/.@cmd}{\setcolor}
\xpatchcmd{\pgfpie@findColor}{\color}{\pgfpie@color}{}{}
\xpatchcmd{\pie}{\color}{\pgfpie@color}{}{}
\makeatother

\begin{document}

\title{Embodiment perception of a smart home assistant}

\author{Mariya Kilina \inst{1}\thanks{The authors contributed equally.\\
Corresponding author \email{mariya.kilina@edu.unige.it}} \and
Tommaso Elia \inst{1}$^*$ \and
Syed Yusha Kareem\inst{1} \and Alessandro Carf\'{i} \inst{1} \and Fulvio Mastrogiovanni \inst{1}}
\authorrunning{M. Kilina et al.}

\institute{Department of Informatics, Bioengineering, Robotics, and Systems Engineering (DIBRIS), University of Genoa, Genoa, Italy}
\maketitle  

\begin{abstract}
Demographic growth and rise in the average age of the population is increasing the demand for the elderly assistance. Health care oriented ambient intelligence technologies are fundamental to support elderly peoples’ autonomy. In this paper, we present a smart home system that is able to recognize human activities and is integrated with a proactive vocal assistant. We chose one of possible user scenarios to show the performance of this smart home system and to perform a preliminary comparison between users’ experience while watching videos of a volunteer interacting with an embodied versus a not-embodied assistant. The scenario is recorded from the user’s point of view, while the user interacts with a robot assistant or a simple vocal assistant. The results of the User Experience Questionnaire show that participants found the robot assistant considerably more attractive, innovative and stimulating in comparison to the vocal assistant.

\end{abstract}

\newpage

%%%%%%%%%%%%%%%%%%%%%%%%%%%%%%%%%%%%%%%%%%%%%%%%%%%%%%%%%%%%%%%
\section{Introduction}
\label{sec:intriduction}

The continuous increase in the age of the population is driving the development of new systems to take care of the elderly \cite{baldissera2018services}. In this context, research on smart homes and Ambient Intelligence (AmI) \cite{sadri2011ambient} is particularly relevant, since the distribution of sensors in the environment, envisioned by AmI, can be used to monitor and support older people at home \cite{chan2008review}. This new paradigm takes the name of Ambient Assisted Living (AAL)\cite{jara2011internet}. AAL systems can monitor various parameters (e.g., weight, heartbeat, human movements) using sensors both distributed in the environment and worn by the user. This information can be processed to monitor the user's health status or to track the user's behaviors. Therefore, such a system may generate relevant information for healthcare professionals \cite{costa2017advances} or for an intelligent system to support the subject \cite{sadri2011ambient}. According to its functionalities, an AAL system can target different applications, such as: i) comfort, i.e., classical home automation allowing the user to control household appliances; ii) health monitoring, i.e., assessment of the user's health status; iii) security, i.e., detection of dangerous situations or emergencies (e.g., fall, smoke, and intrusion detection).

In healthcare AAL applications, raw data can be provided to caregivers to monitor the person's health, or it can be processed to extract higher-level information. An example of this process is the recognition of activities of daily living (ADL) such as: walking, drinking, and brushing teeth \cite{bruno2013analysis,ruzzon2020multi}. The interest of healthcare professionals in automatic ADL recognition derives from their relationship with the degree of independence of the subject \cite{gama2000association}. In addition, an AAL system that can recognize ADLs can use this information to suggest activities (e.g., encouraging exercise after sitting for a long time) or remind people to do things (e.g., reminding them to take medication before or after a meal). These applications present different challenges in terms of sensors to be used, data management, data analysis, and user interaction \cite{ranasinghe2016review}. 

Additionally, the proper ADL recognition method should be selected. Most of the ADL recognition systems in the literature analyze human motion to determine the corresponding ADL \cite{lin2018activity}. However, solutions that consider context have also been proposed since movements may have a different meaning depending on the circumstances \cite{kareem2018arianna}. Finally, the system should interact in a natural and non-invasive way to inspire the user's trust. Nowadays, there are many alternatives for an intelligent system to interact with a human. The most common interfaces use voice feedback or visual stimuli on a screen. At the same time, the emerging field of social robotics proposes to use robots as trusted partners for humans in a home or hospital scenarios. 

Previous work in the field of social robotics show that some features of the intelligent system such as perceived sociability, animacy and humanlike-fit can predict the user's willingness to use the system \cite{wagner2019human}. Thereby, users appreciate intelligent systems with which they can communicate using natural languages and which behave like humans. In this context, the agent embodiment is an important aspect influencing user's perception of the whole system. In Virtual Reality (VR) environment users prefer visible assistant rather then a voice one \cite{kim2019effects, kim2018does}. Moreover, an assistant that is able to move around the space and interact with objects is perceived as more influential, trustworthy and socially present compared to the embodied, but motionless one \cite{kim2018does}. Speaking of the comparison in perception between smart speaker, i.e., vocal assistant and robot assistant, similar results are observed. Vocal assistants are perceived as more socially present if they have human-like embodiment \cite{kontogiorgos2019effects, nakanishi2020smart}. Furthermore, users tend to trigger conversations with a robot assistant significantly more often then with a smart speaker, i.e., vocal assistant \cite{nakanishi2020smart}. All of these findings lead us to the assumption, that along with the vocal assistants, AAL systems could benefit from the human-like embodiment of their agents in terms of the user experience.

Thus, the goal of this work is to investigate if the type of physical embodiment of a virtual assistant has a positive effect on users of an AAL system. To this end, we integrated a voice reminder assistant into a state-of-the-art ADL recognition system, and we recorded a video with two experimental conditions: i) a voice interface with a speaker and ii) a voice interface mediated by a small social robot (i.e., SOTA). The videos were provided to a set of volunteers who were asked to answer a user experience questionnaire after watching a video. The remainder of the article follows. Section 2 presents the overall system's architecture. Section 3 explains the experimental setup and the assessment method. Section 4 presents the results of the experiment. Finally, Section 5 present conclusions.

%%%%%%%%%%%%%%%%%%%%%%%%%%%%%%%%%%%%%%%%%%%%%%%%%%%%%%%%%%%%%%%
\section{System's Architecture}
\label{sec:system_architecture}

\begin{figure}[t]
	\centering
	\includegraphics[width=0.8\textwidth]{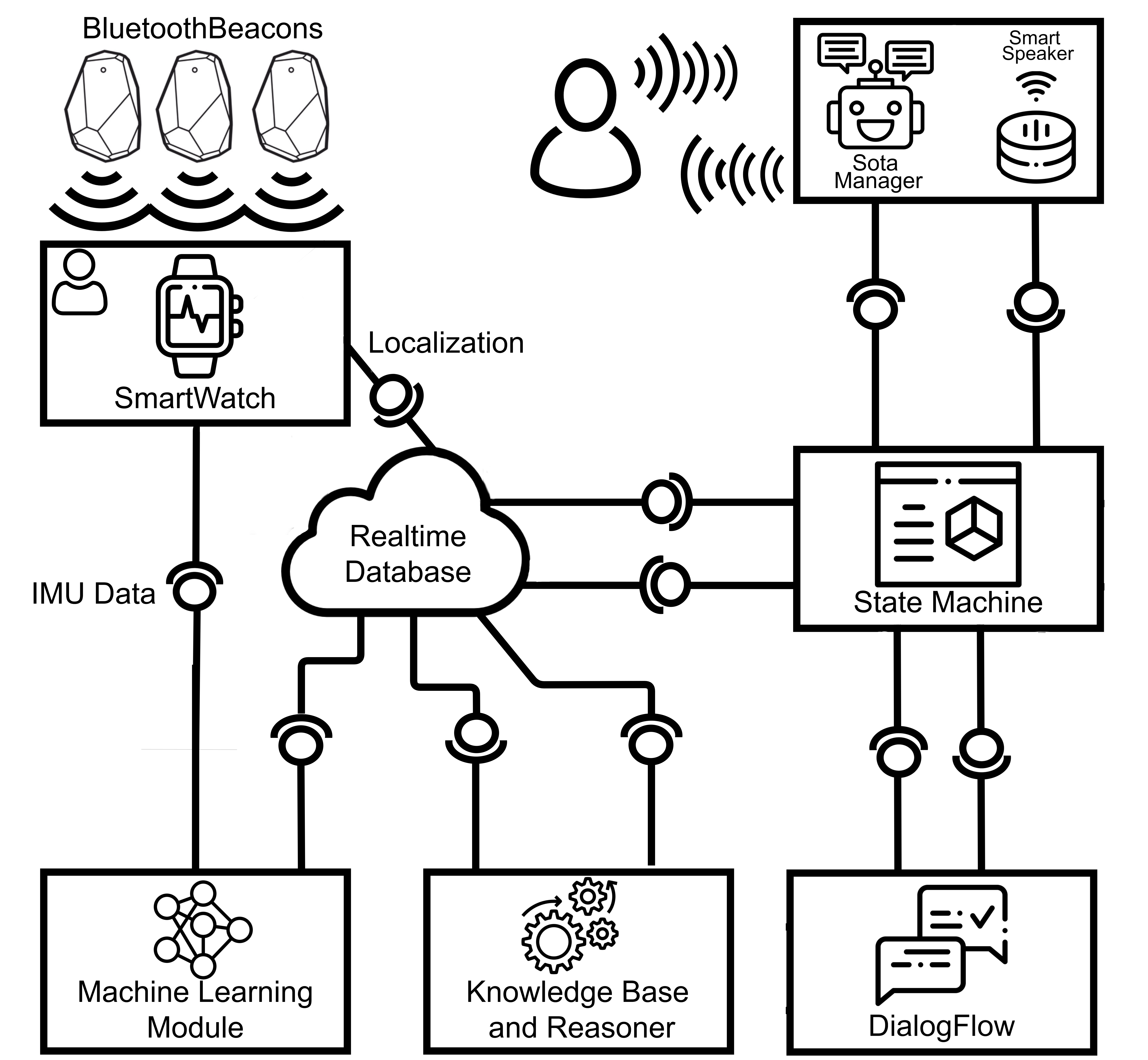}
	\centering \caption{System's architecture.}
	\label{fig:software_architecture}
\end{figure}

In this section we broadly describe the architecture of the system. For more detailed information about all of it's components refer to Kareem et al., 2018 \cite{kareem2018arianna}.

As shown in Figure~\ref{fig:software_architecture}, the system consists of several interconnected parts. Specifically, the system is composed of three macro-layers connected to each other via a database:

\begin{itemize}
    \item[] \textbf{Sensing Layer:} acquires sensory data coming from a smart watch.
    \item[] \textbf{Human Activity Recognition (HAR) Layer:} models and recognizes activities of daily living (ADL).
    \item[] \textbf{Human Computer Interaction (HCI) Layer:} provides interfaces for human computer interaction.
\end{itemize}

%%%%%%%%%%%%%%%%%%%%%%%%%%%%%%%
\subsection{Sensing Layer}
\label{subsec:sensing_layer}

In the Sensing Layer sensory data is collected by using a smart watch and a set of Estimote Beacons~\footnote{https://estimote.com/} placed in each room of the user's house. This system allows to trace the user's location based on which beacon the smart watch is connected to. This information is acquired and then published to the database by an Android application, installed on the smart watch. Another application acquires the inertial raw data from the smart watch and sends it to the Machine Learning module that recognizes simple human activities. 
 
%%%%%%%%%%%%%%%%%%%%%%%%%%%%%%%
\subsection{Human Activity Recognition Layer}
\label{subsec:human_activity_recognition_layer}

We take a hybrid approach towards HAR. On the one hand, adopting a machine learning approach for recognizing simple activities based on the IMU data stream from the smartwatch, e.g., (i) drinking, (ii) pouring, (iii) standing up, (iv) sitting down, and (v) walking. On the other hand, adopting an ontology based approach for recognizing contextual activities, e.g., (i) having breakfast and (ii) sitting idle for a specific amount of time. To be able to use the ontology as a part of overall HAR system's architecture, we used the OWLOOP API~\footnote{https://github.com/TheEngineRoom-UniGe/OWLOOP} that allows adding, removing and reasoning over axioms present in the ontology from OOP (object oriented programming) domain. 

Machine learning module and Arianna \cite{kareem2018arianna} (i.e., ontology and reasoner) interact with each other via a real-time database. As soon as there is new information in the real-time database, regarding the user's location and/or the simple activity performed, Arianna uses that information to make assertions in the ontology and reason based on those assertions. The reasoner we used was Pellet. If reasoning leads to the inference of a contextual activity, then this information is saved back to the real-time database. Lastly, the human computer interaction layer becomes active.

%%%%%%%%%%%%%%%%%%%%%%%%%%%%%%%
\subsection{Human Computer Interaction Layer}
\label{subsec:human_computer_interaction_layer}

The system interacts with the assisted person through a vocal interface. We introduced two different ways of interaction: (i) through a vocal assistant (i.e., a smart speaker) and (ii) through a robot assistant (i.e., a small social robot named Sota). Depending on the context (i.e., user's location, user's contextual activities, knowledge from the database), Arianna is able to trigger a specific conversation. The system reacts to the continuous database update online and manages the dialogue with the user through the Dialogflow SDK used to define the logical flow of the dialogues. 

%%%%%%%%%%%%%%%%%%%%%%%%%%%%%%%
% \subsubsection{Sota manager and Smart speaker}

\begin{figure}[t]
	\centering
	\includegraphics[width=12.1cm]{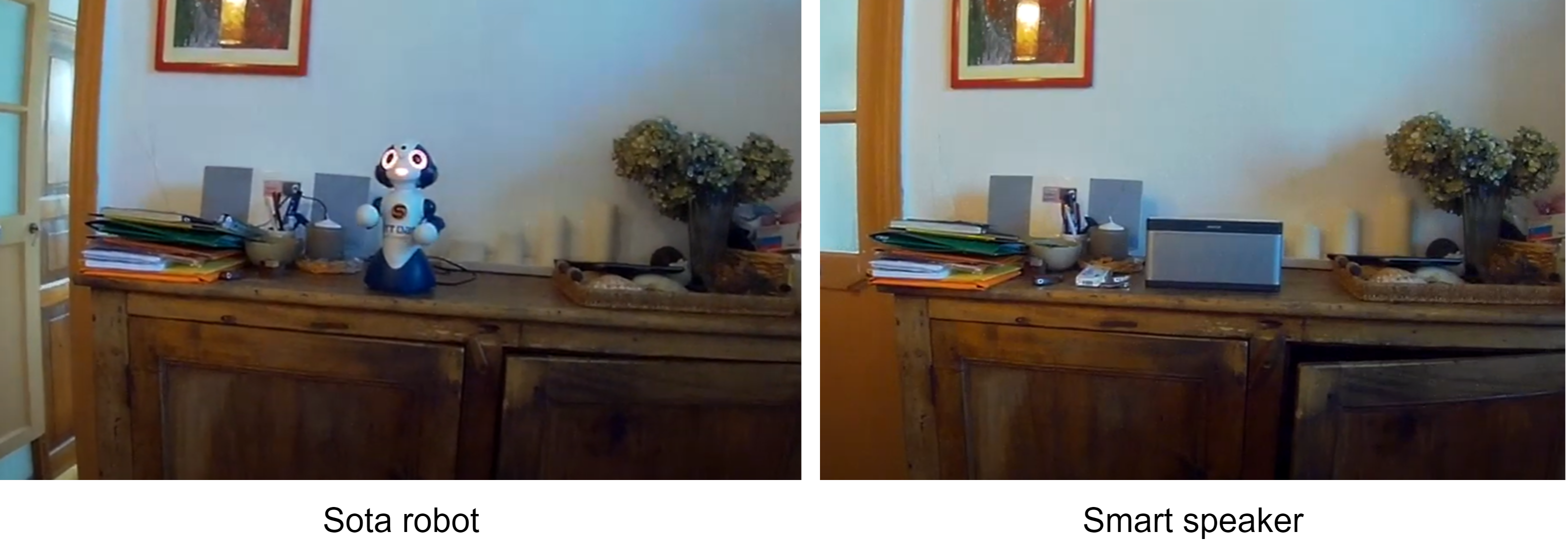}
	\centering \caption{The robot assistant (Sota robot) on the left and the vocal assistant (smart speaker) on the right.}
	\label{fig:robotAndVocalAssistant}
\end{figure}

Figure~\ref{fig:robotAndVocalAssistant} shows the robot and vocal assistants. Both the Sota manager module and the Smart speaker module are connected to the real time database and Dialogflow and both, based on the context provided by the real-time database, use the Dialogflow SDK to handle the microphone and the speaker signals. The difference between the robot and speaker conditions is that in the robot condition Sota robot can perform gestures and movements while conversing with the user.

%%%%%%%%%%%%%%%%%%%%%%%%%%%%%%%%%%%%%%%%%%%%%%%%%%%%%%%%%%%%%%%
\section{The Experiment}
\label{sec:the_experiments}

\subsection{Experimental setup}
\label{sec:experimental_setup}

To evaluate the system, we focused on deploying a use case demonstrating the potential of context-based proactive human-computer interaction. The use case is designed for elderly individuals living independently and is oriented towards their support and well-being. Specifically, for our experiment, we developed a use case where the system reminds the user about the medications they have to take.

The use case scenario is as follows. Arianna detects the user's presence in the kitchen and recognizes the breakfast activity. Arianna knows that the user must take particular medicines in the morning on a full stomach. Therefore, Arianna reminds the user to take the required medications with a glass of water. To do this, Arianna uses the vocal assistant located on the kitchen table and tries to recognize whether the user takes the medications or not (e.g., by detecting whether the user has accessed the medicine cabinet or if the user pours and drinks water). After some time, Arianna also asks the user if they took the medications.

\begin{figure}[t]
	\centering
	\begin{subfigure}[t]{0.6\textwidth}
	    \includegraphics[width=\textwidth]{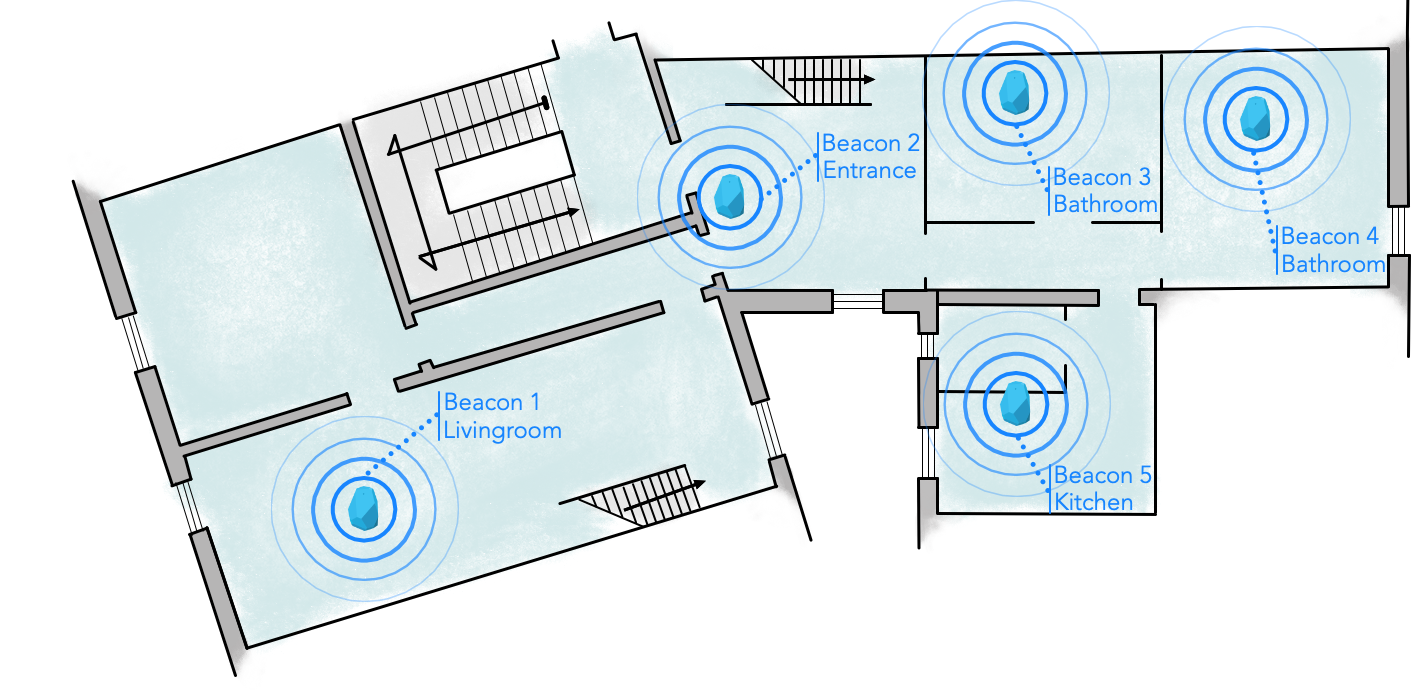}
        \caption{First floor.}
        \label{fig:firstFloor}
	\end{subfigure}
	\begin{subfigure}[t]{0.35\textwidth}
		\includegraphics[width=\textwidth]{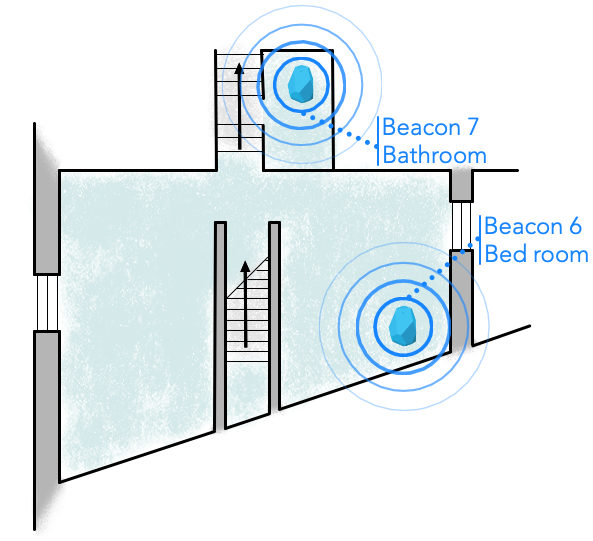}
		\caption{Second floor.}
		\label{fig:secondFloor}
    \end{subfigure}
\caption{Location of Estimote beacons on the map apartment.}%
\label{fig:EstimoteFloor}	
\end{figure}

To set up the experimental environment, we installed the Estimote beacons in a real home. The location of all beacons placed in the house is shown in Figure~\ref{fig:EstimoteFloor}. Since the Estimote beacons interfere with each other, proper proximity-based user localization requires beacons' calibration. The calibration process includes placing the beacons strategically and tuning the Bluetooth range of each beacon. After that, with the support of a volunteer wearing a smart watch, we performed experiments and tested the use case.

\begin{figure}[t]
\captionsetup[subfigure]{labelformat=empty}
	\centering
	\begin{subfigure}[t]{0.3\textwidth}
	 \includegraphics[width=\textwidth]{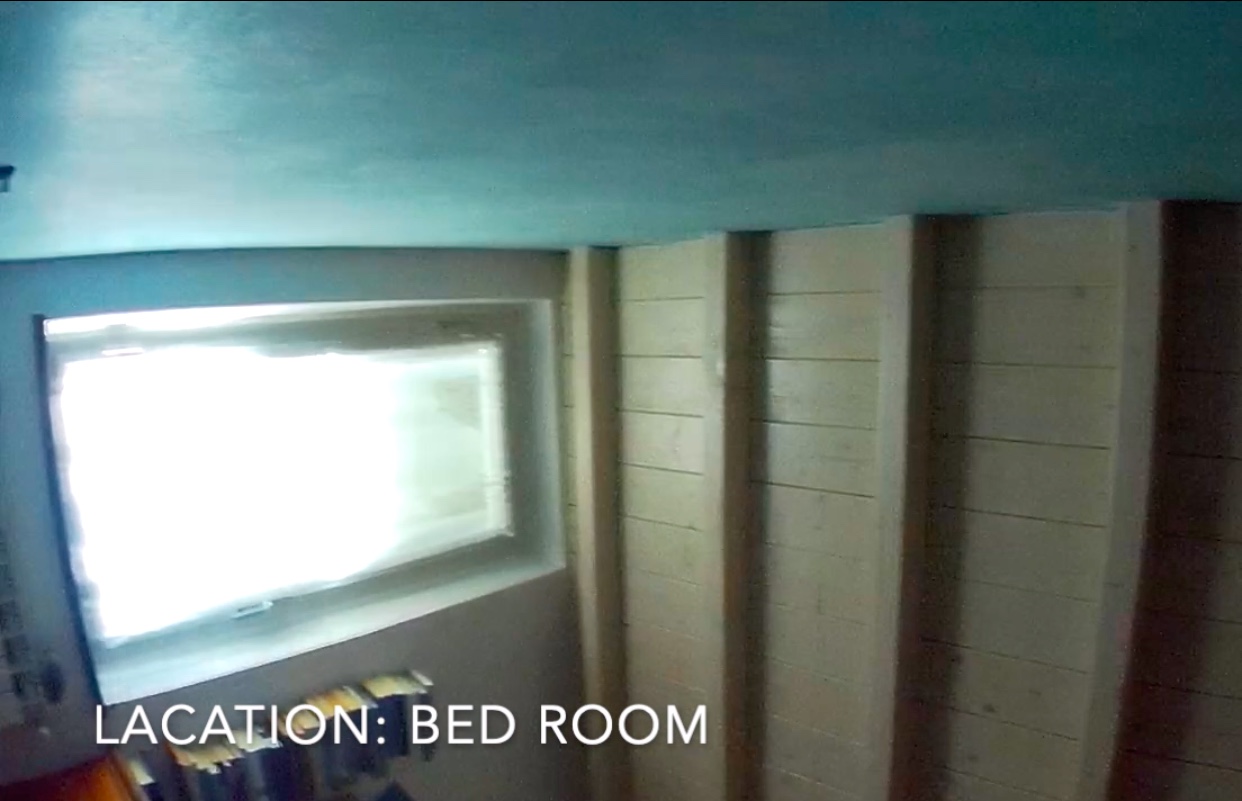}
     \caption{(i) In the bedroom}
	\end{subfigure}
	%\begin{subfigure}[t]{0.24\textwidth}
	%	\includegraphics[width=\textwidth]{images/sequenceSecondUseCase/2.jpg}
	%	\caption{(ii) Getting out of bed}
	%\end{subfigure}
	\begin{subfigure}[t]{0.3\textwidth}
		\includegraphics[width=\textwidth]{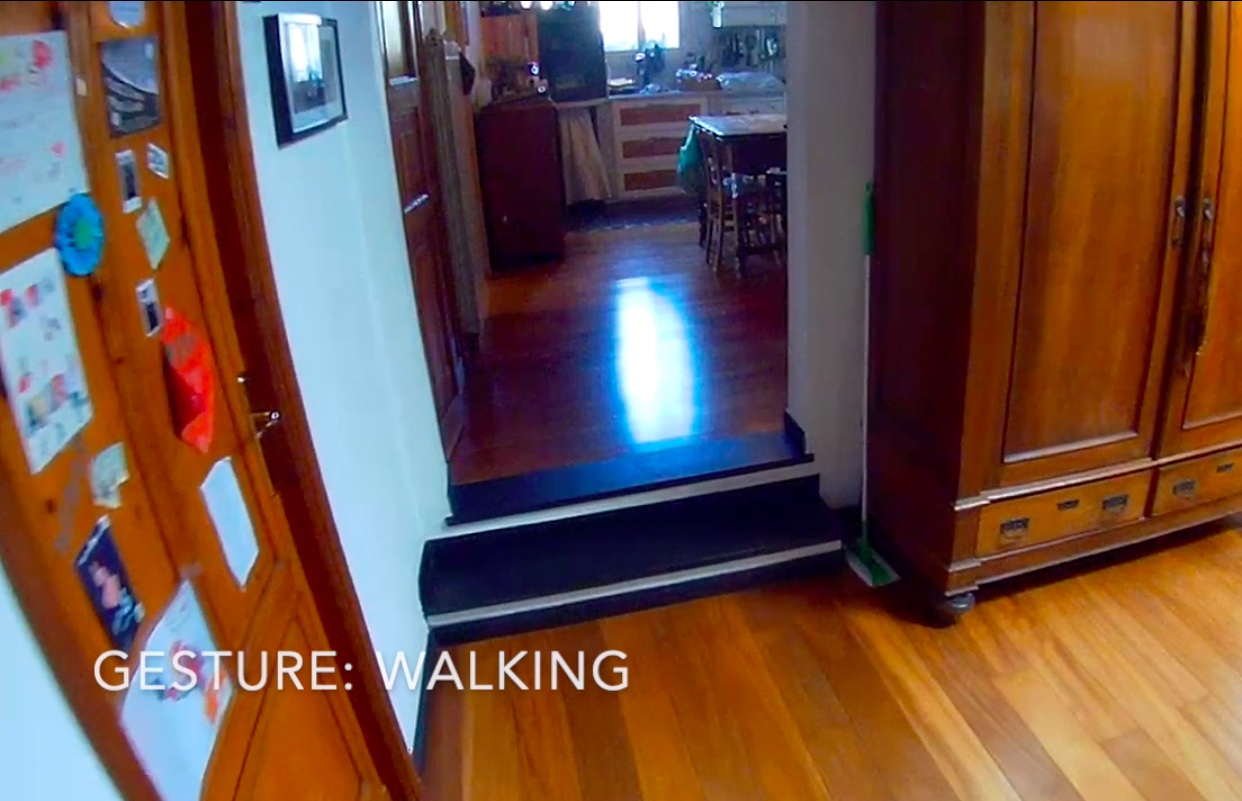}
		\caption{(ii) Walking }
	\end{subfigure}
	%\begin{subfigure}[t]{0.24\textwidth}
	%	\includegraphics[width=\textwidth]{images/sequenceSecondUseCase/4.jpg}
	%	\caption{(iii) Entering the kitchen} 
	%\end{subfigure}
	\begin{subfigure}[t]{0.3\textwidth}
		\includegraphics[width=\textwidth]{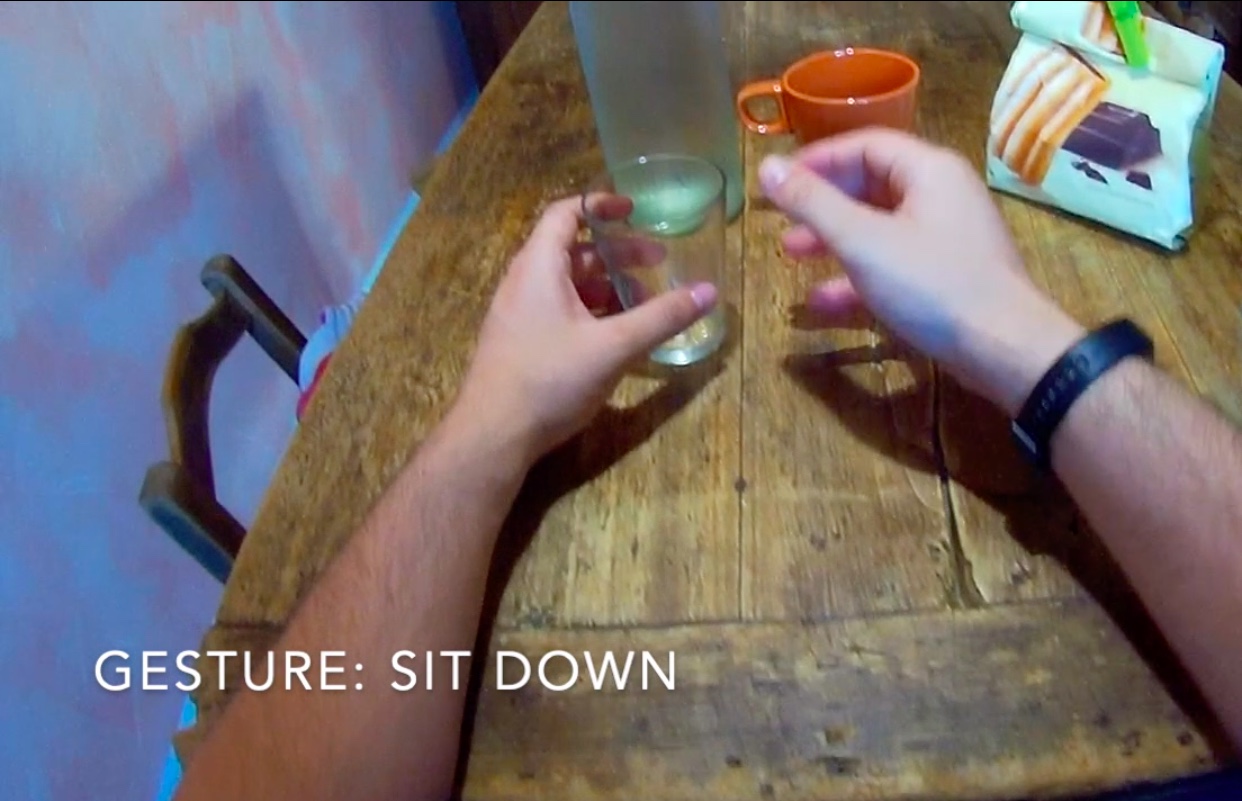}
		\caption{(iii) Sitting at the table}
	\end{subfigure}
	\par\medskip
	\begin{subfigure}[t]{0.3\textwidth}
		\includegraphics[width=\textwidth]{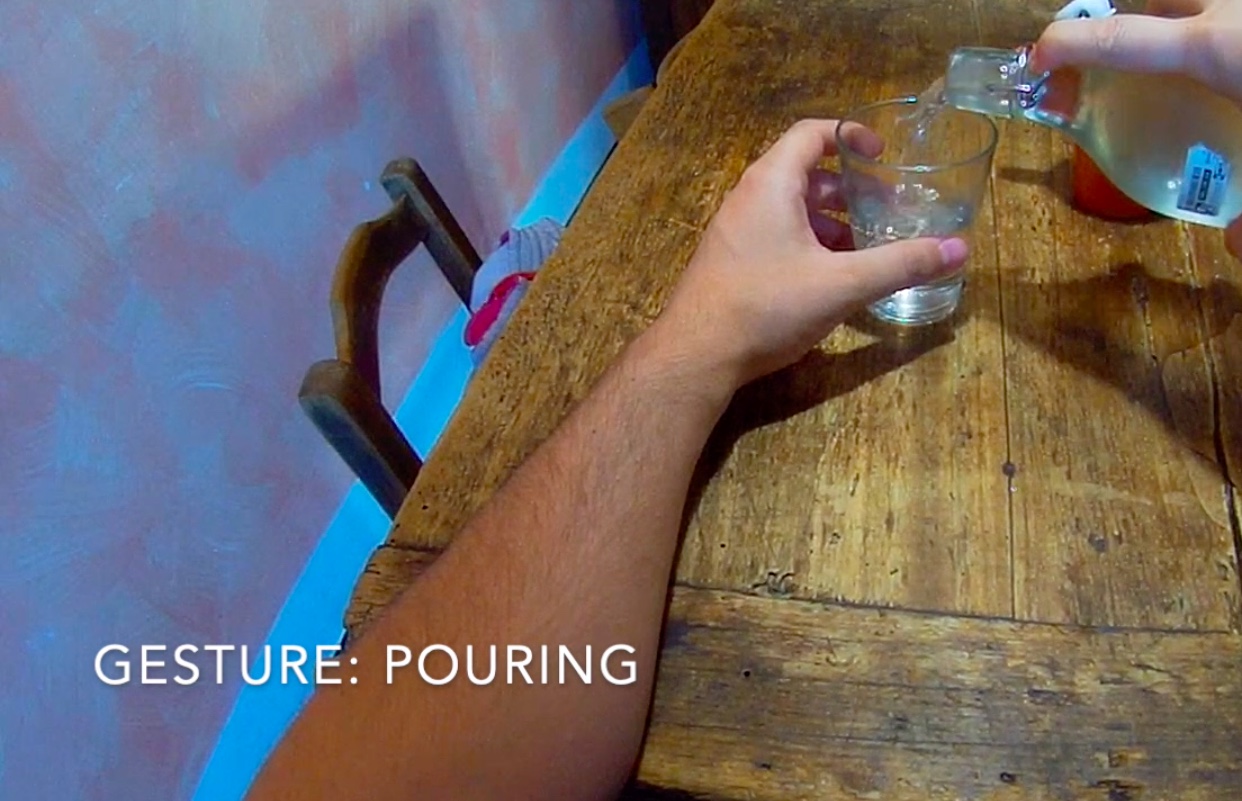}
		\caption{(iv) Pouring water }
	\end{subfigure}
	\begin{subfigure}[t]{0.3\textwidth}
		\includegraphics[width=\textwidth]{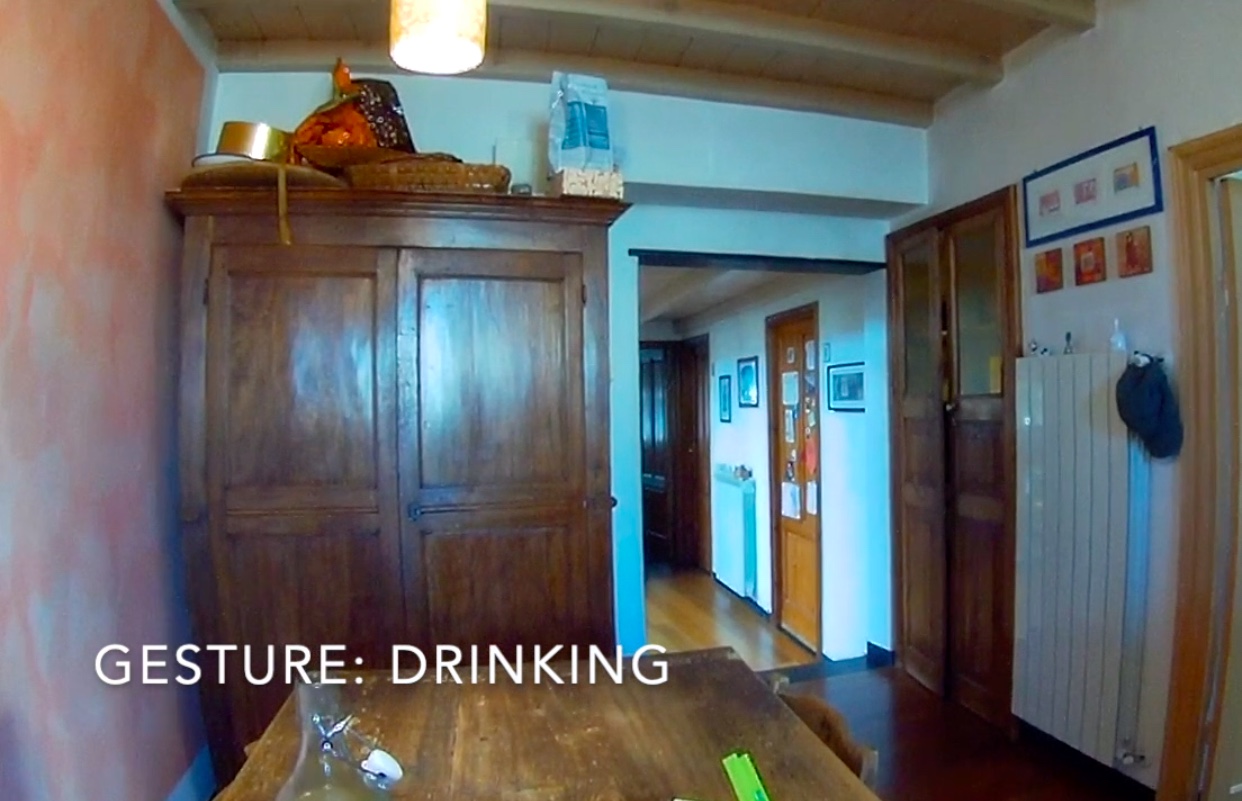}
		\caption{(v) Drinking}
	\end{subfigure}
	\begin{subfigure}[t]{0.3\textwidth}
		\includegraphics[width=\textwidth]{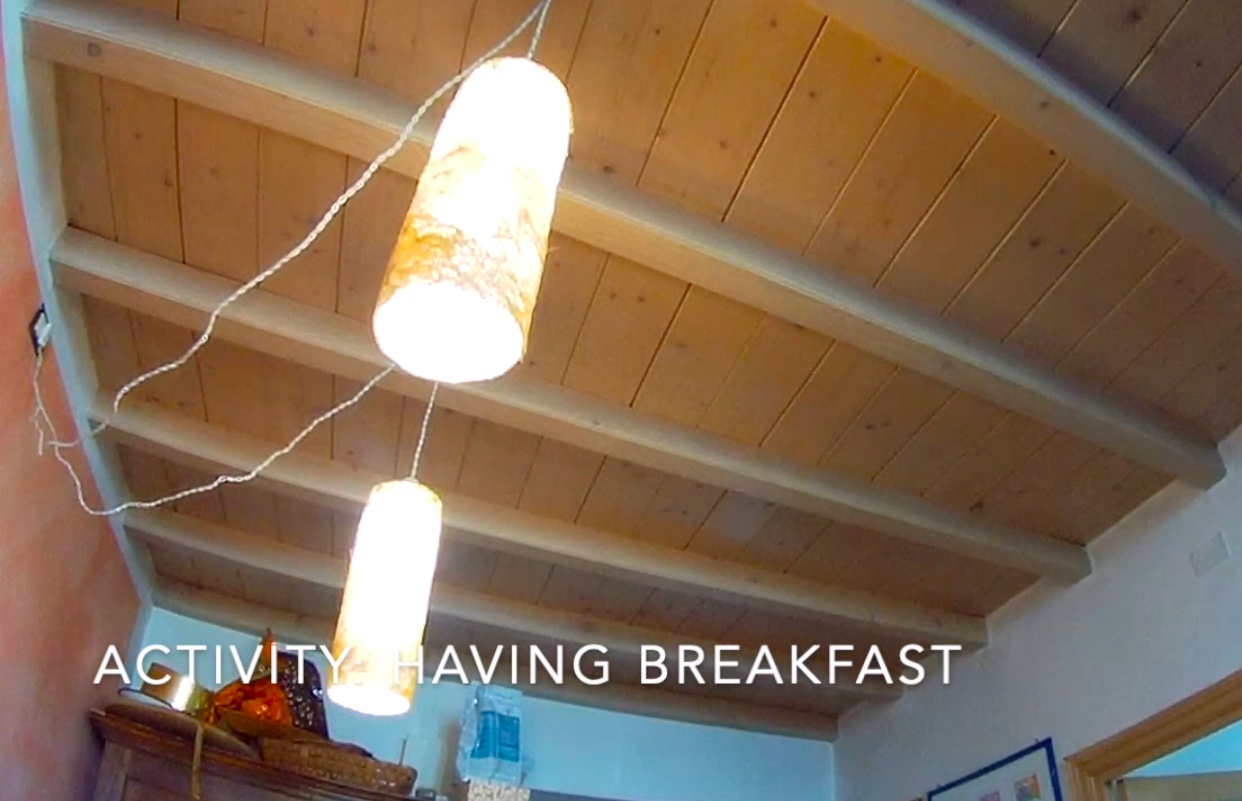}
		\caption{(vi) Having breakfast}
	\end{subfigure}
\caption{Snapshots from the use case recording.}
\label{fig:secondScenario}	
\end{figure}

We recorded a volunteer's interaction with Arianna from an egocentric point of view (i.e., the camera is placed on the volunteer's forehead) as shown in Figure~\ref{fig:secondScenario} with two types of interfaces: a simple smart speaker as the vocal assistant \footnote{https://youtu.be/85BQhc87pqA} and Sota robot as the robot assistant \footnote{https://youtu.be/w9-w5tZRZDE}. We chose to evaluate user experience (UX) by providing participants with a virtual experience of the medication reminder use case. This way, participants watched two videos where a volunteer was interacting (i) with the vocal assistant and (ii) with the robot assistant. At the end of each video participants were asked to fill the User Experience Questionnaire (UEQ) \cite{schrepp2017construction} that we will describe in the following section. All the survey was held online, using Google Forms.

%%%%%%%%%%%%%%%%%%%%%%%%%%%%%%%%%%%%%%%%%%%%%%%%%%%%%%%%%%%%%%%
\subsection{Assessment method}
\label{sec:assessment_method}

For the evaluation of UX of the system we chose the User Experience Questionnaire (UEQ) that was created in Germany in 2005 \cite{schrepp2017construction}. The UEQ is available and validated in 21 languages, including English and Italian. 

The UEQ contains six scales that are listed below, with 26 items in total:
\begin{itemize}
    \item[] \emph{Attractiveness}: Overall impression of the system. Do users like or dislike the system?
    \item[] \emph{Perspicuity}: Is it easy to get familiar with the system? Is it easy to learn how to use the system?
    \item[] \emph{Efficiency}: Can users solve their tasks without unnecessary effort?
    \item[] \emph{Dependability}: Does the user feel in control of the interaction?
    \item[] \emph{Stimulation}: Is it exciting and motivating to use the system?
    \item[] \emph{Novelty}: Is the system innovative and creative? Does the system catch the interest of users?
\end{itemize}

The Attractiveness scale contains 6 items, while other scales contain 4 items each. Items consist of two adjectives that are antonyms to each other, and a seven-stage scale, where +3 represents the most positive response and -3 represents the most negative one. Within each item, the order of the positive and the negative adjectives are randomized. 

Each participant was asked to fill the questionnaire twice: to evaluate their perception of the vocal assistant and of the robot assistant. We randomized the order of questionnaires for these two conditions to avoid introducing a bias. Half of the participants received a questionnaire with the vocal assistant first and the robot assistant next (Q1), another half received it in a reversed order (Q2). In addition to evaluating the system, we asked participants to specify some general information about themselves, such as their age, experience in a scientific/technological sector and the evaluation of their previous uses of a technology similar to the proposed system. 

\subsection{Participants}
\label{sec:participants}

The surveys has been shared and compiled by 240 volunteers (134 males, 103 females, 3 decided not to answer), between 12 and 69 years old (median = 29.5, STD = 14.14). More than a half of them (52.5\%) worked in a scientific/technological sector. 64.6\% of the participants used a vocal assistant at least once (16.8\% of which use it every day and 37.4\% several times a week), and 14.6\% used a robot assistant (14.3\% of them use it every day and 80\% use it once a month or less).

%%%%%%%%%%%%%%%%%%%%%%%%%%%%%%%%%%%%%%%%%%%%%%%%%%%%%%%%%%%%%%%
\section{Results}
\label{sec:results}

First, we verified that the order of the two versions of the system (with robot assistant and with vocal assistant) does not introduce a bias. To do that, we compared Q1 and Q2 questionnaires over all of the UEQ scales. We used the T-Test to check if the scale means differ significantly. As suggested in the literature, Alpha-Level 0.05 was used \cite{edgell1984effect}. The results showed Alpha-Level higher then 0.05 in all the comparisons. Thus, no significant difference was noticed, meaning that the evaluations of the two types of the system are not influenced by the order in which they were presented.

\begin{figure}[t]
    \centering
    \includegraphics[width=0.98\textwidth]{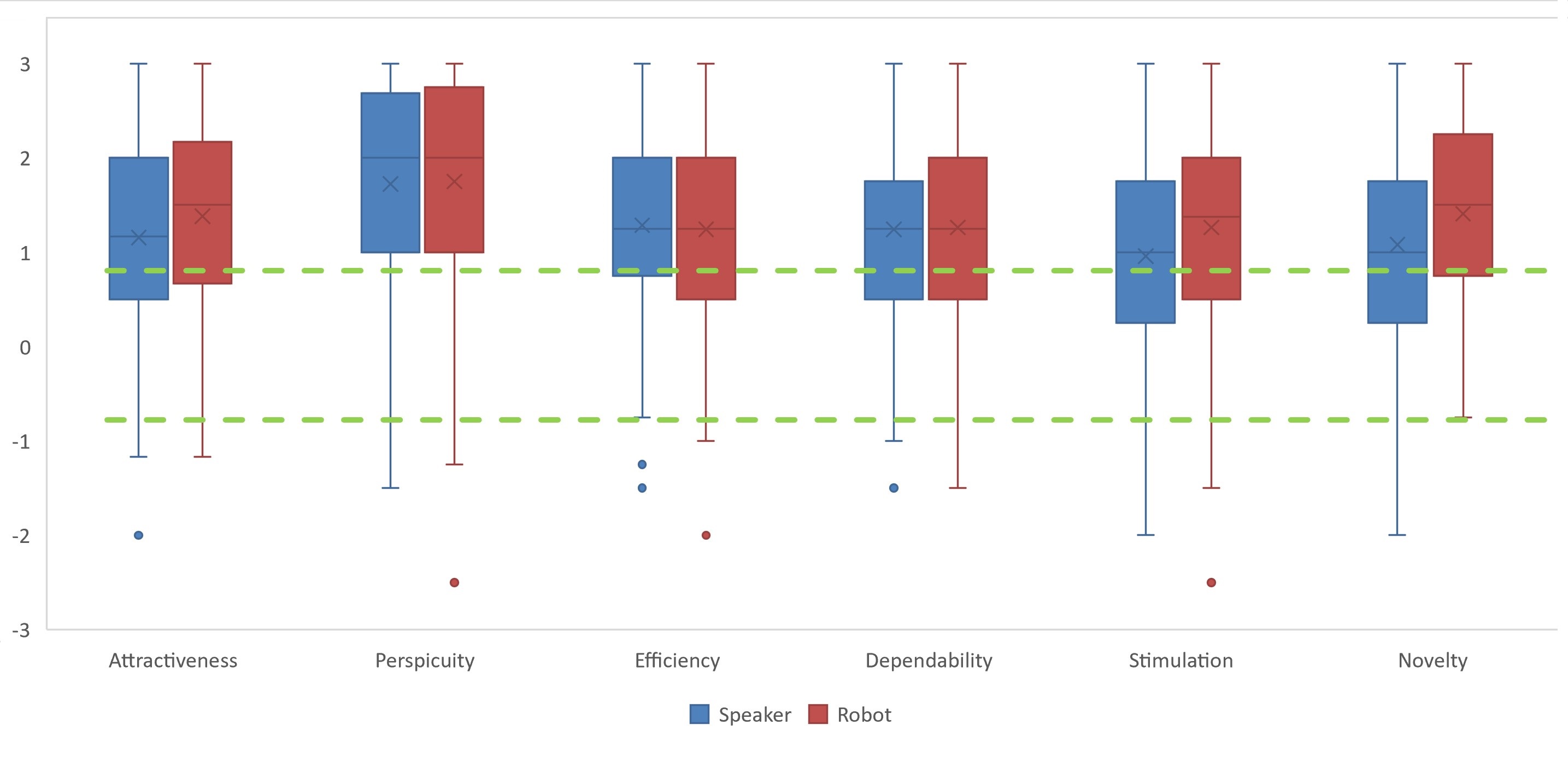}
    \caption{Comparison between users' perception after interaction with the robot assistant and the vocal assistant.}
    \label{fig:comparison-NEVI-EVI}
 \end{figure}

Next, we evaluated both systems with the vocal assistant and the robot assistant individually and then compared them. We analyzed the system equipped with the vocal assistant. In the Figure~\ref{fig:comparison-NEVI-EVI} we provide the means and the variance for each scale evaluated. Values between $-0.8$ and $0.8$ represent neutral evaluation of the corresponding scale, values higher then $0.8$ represent positive evaluation and values smaller then $-0.8$ represent negative evaluation. Then, the same considerations were made for the system equipped with the robot assistant.

\newcommand*{\redbullet}{\raisebox{-.5ex}{\textcolor[RGB]{192,80,78}{\Huge \textbullet}}}
\newcommand*{\bluebullet}{\raisebox{-.5ex}{\textcolor[RGB]{79, 129, 188}{\Huge \textbullet}}}

As we can see from the Figure~\ref{fig:comparison-NEVI-EVI}, the results are promising. The means for all the scales are far above the threshold of $0.8$, and we can also notice a high value of \emph{Perspicuity} that tells us that the system is easy to get familiar with and use.  

The two systems had success among the volunteers. Nevertheless, we evaluated which of the two systems is considered the preferred one. From Figure~\ref{fig:comparison-NEVI-EVI}, it is evident that the interaction with a robot assistant always has a higher or equal mean. This suggests that the volunteers preferred the interaction with the robot assistant rather than the interaction with the vocal assistant. This was further confirmed with the Wilcox test.

From the Wilcox test we can note that the mean difference is significant for 3 of the 6 scales of interest: attractiveness ($p < 0.01$), stimulation ($p < 0.001$) and novelty ($p < 0.001$). Therefore, in our study the robot assistant has been perceived as more attractive, stimulating and innovative than the vocal assistant.

\section{Conclusion}
\label{sec:conclusion}
Nowadays, it is fundamental to extend the autonomy of the elderly population as much as possible. Through a distributed sensor system (i.e., smart home), it is possible to track human activities, guaranteeing the autonomy of the subject while monitoring their health state. However, by their nature, the limit of these types of technologies is the perception of intrusiveness. To cope with this limitation, this research proposes a solution that, in addition to simply monitoring human activities, uses user information to offer an intuitive and engaging interactive user experience allowing users to perceive the solution as more attractive.
For this purpose, a proactive vocal assistant and a proactive robot assistant have been integrated into an AAL system.

Based on the results shown in section \ref{sec:results}, this research work proposes an alternative vision to the classic smart home systems for human activities monitoring. In fact, we observed that by integrating a user-friendly interface elements such as voice or robot assistants, we can favor a more positive user experience. What emerges from the results is that (i) in terms of functionality, both robot and vocal assistant have been perceived positively (Figure~\ref{fig:comparison-NEVI-EVI}), (ii) whereas, in terms of \textit{attractiveness, stimulation and novelty}, the robot assistant has better results compared to vocal assistant (Figure \ref{fig:comparison-NEVI-EVI}). In summary, even if both experiences with the vocal and the robot assistant are positively perceived by the users, the robot assistant, because of its nature, is perceived as more attractive, stimulating and novel.

Although these results are supported by previous literature on agent embodiment, this research field should be the subject of additional studies. In particular, carrying on studies where subjects experience first-hand different agents' embodiment is fundamental.

\bibliographystyle{splncs04}
\bibliography{bibl}

\end{document}